# THE HELIUM–RICH CATACLYSMIC VARIABLE

# ES CETI

CATHERINE ESPAILLAT,[1,2] JOSEPH PATTERSON,[2] BRIAN WARNER,[3] & PATRICK WOUDT[3]

[1] Department of Astronomy, University of Michigan, 830 Dennison Building, Ann Arbor, MI 48109; ccespa@umich.edu

[2] Department of Astronomy, Columbia University, 550 West 120th Street, New York, NY 10027; jop@astro.columbia.edu

[3] Department of Astronomy, University of Cape Town, Rondebosch 7700, South Africa; warner@physci.uct.ac.za, pwoudt@artemisia.ast.uct.ac.za

## ABSTRACT

We report photometry of the helium-rich cataclysmic variable ES Ceti during 2001–2004. The star is roughly stable at *V*~17.0 and has a light curve dominated by a single period of 620 s, which remains measurably constant over the 3 year baseline. The weight of evidence suggests that this is the true orbital period of the underlying binary, not a "superhump" as initially assumed. We report GALEX ultraviolet magnitudes, which establish a very blue flux distribution ($F_\nu \sim \nu^{1.3}$), and therefore a large bolometric correction. Other evidence (the very strong He II 4686 emission, and a ROSAT detection in soft X-rays) also indicates a strong EUV source, and comparison to helium-atmosphere models suggests a temperature of 130±10 kK. For a distance of 350 pc, we estimate a luminosity of $(0.8–1.7)\times10^{34}$ erg/s, yielding a mass accretion rate of $(2–4)\times10^{-9}$ $M_\odot$/yr onto an assumed 0.7 $M_\odot$ white dwarf. This appears to be about as expected for white dwarfs orbiting each other in a 10 minute binary, assuming that mass transfer is powered by gravitational radiation losses. We estimate mean accretion rates for other helium-rich cataclysmic variables, and find that they also follow the expected $\dot{M} \sim P_o^{-5}$ relation. There is some evidence (the lack of superhumps, and the small apparent size of the luminous region) that the mass transfer stream in ES Cet directly strikes the white dwarf, rather than circularizing to form an accretion disk.

*Subject headings*: accretion, accretion disks — binaries: close — novae, cataclysmic variables — stars: individual (ES Ceti, KUV 01584–0939)

# 1. INTRODUCTION

The star KUV 01584–0939 (hereafter ES Ceti) was discovered as an ultraviolet-bright object during the Kiso survey (Kondo, Noguchi, & Maehara 1984), with strong He II and weak Balmer emission (Wegner, McMahan, & Boley 1987). It soon became clear that the "Balmer" lines are really He II Pickering lines, which identifies ES Cet as essentially a pure helium star. Warner & Woudt (2002, hereafter WW) found a persistent photometric period of 620.2 s, certifying ES Cet as a member of the AM Canum Venaticorum class: cataclysmic variables containing a pair of helium white dwarfs.

Since they possess Roche-lobe-filling white dwarfs, all stars in this class have a very short orbital period; and their light curves are typically dominated by variations at the orbital period $P_o$ and a closely spaced "superhump" period. Superhump variations, which are usually dominant, arise from the precession of the accretion disk (Whitehurst 1988, Hirose & Osaki 1990). These periods can be quite difficult to disentangle and interpret; confusion between them has been a significant barrier to progress. Therefore we have conducted a three-year study of ES Cet, in order to parse the periodic signal into its correct components. Here we report details of that study, along with a general study of the AM CVn class.

# 2. OBSERVATIONS AND LIGHT CURVES

The optical observations consist of nightly time-series photometry obtained with the 1.3m McGraw-Hill telescope at the MDM Observatory on Kitt Peak, and several telescopes at the South African Astronomical Observatory. The observations covered ~140 hr over 51 nights during 2001–4, and are summarized in Table 1. The data are differential CCD photometry (Skillman & Patterson 1993, O'Donoghue 1995), usually with a wide bandpass (~4000–8000 Å) to maximize throughput. ES Cet's mean brightness was roughly stable (near $V$=17.0) from night to night, with an average peak-to-peak range of ~0.12 mag, as seen in the raw light curve of Figure 1. This was also true in the WW data. The nightly power spectra were also essentially identical to those of WW: a fundamental frequency of 139.3 c/d and several harmonics.

# 3. PERIOD ANALYSIS

In order to study fine structure in the periodicity, we selected our densest data streams and calculated separate power spectra. The best interval was in 2002 October (JD 2452550–6), and the relevant portions of that power spectrum are shown in Figure 2. The star exhibits, to a first approximation, a fairly pure signal at 139.31(1) c/d, with harmonics but no obvious sideband structure. Analysis of the other dense data streams (2001 October and 2002 September) gave a similar result.

These three time series are long enough to yield precise mean waveforms for the 620 s signal, and these are shown in Figure 3. Maxima and minima are well defined and the waveform is essentially stable. Individual-night waveforms also agreed with the general pattern (slow rise, fast decline). In order to track the stability of the 139.31 c/d signal, we constructed an O–C diagram using the nightly timings of minimum light (Table 2). With respect to a suitably chosen test ephemeris with constant period, the O–C residuals essentially trace out a straight line of zero

slope — indicating the constancy of the period. This is shown in Figure 4. More precisely, the star tracks the ephemeris

$$\text{Minimum light} = \text{HJD } 2452201.3941(2) + 0.007178376(2)\ E. \qquad (1)$$

The $2\sigma$ upper limit on $\dot{P}$ is $|\dot{P}|<1.5\times10^{-11}$.

Having ascertained that the signal is of good frequency and amplitude stability, we decided to refine the search for fine structure by "cleaning" the time series for strong signals. We subtracted the obvious signals, calculated the power spectrum of the residuals, and then re-inserted the strong signals (which tower off-scale) in the power spectrum. The relevant results are seen in Figure 5, showing only a broad bump on the high-frequency side of the strong signals.

The individual peaks comprising these broad bumps correspond to very low Fourier amplitudes — all less than 0.01 mag. For our time series, these are too weak to permit a secure frequency measurement. So the main result here is failure; we could not certify and measure any fine structure. But since the windowing of our time series is far from ideal, and since such asymmetry (blue shift) is unusual in our experience, the bumps deserve a mention and might be worth further study. They all sit ~12 c/d displaced from the main signal, and their asymmetry about the main signal assures that they are not an artifact of simple amplitude modulation (which produces equal power on both sides).

## 4. THE AM CVN CREDENTIALS

ES Cet's helium-rich spectrum, very short period, and constant magnitude certify it as a very likely member of the AM Canum Venaticorum class of CVs. The very blue colors suggest a good match to HP Librae and AM CVn itself. But for the latter two stars, and indeed for all previous members of the AM CVn class, the dominant periodic signal is the superhump (see e.g. O'Donoghue 1995). That is what we expected when we began this study; but there is precious little evidence to support this.

Precise parsing of the signals has been done for just two stars in this class: AM CVn (Skillman et al. 1999) and HP Lib (Patterson et al. 2002). These stars show orbital variations of ~0.012 mag full amplitude, and superhump variations of 0.02–0.05 mag. The orbital periods are stable over the longest measurable baselines (years), while superhump periods wander slightly on timescales of months. The superhump harmonics are also characterized by a subtle but regular sideband structure (see Figures 3–4 and Table 2 of Skillman et al. 1999).

In the ES Cet time series we searched carefully for both of these superhump earmarks, and did not find them. Figures 3 and 4 demonstrate the stability of the periodic signal. Figure 5 suggests that there may be some fine-structure near the strong signals; but unlike the case of other AM CVn stars, we could not find any precise frequency shift associated with these weak bumps. Thus the fine structure test also fails.

Excuses might be found for such malfeasance, but the most likely hypothesis is simply that the photometric period is the *orbital* period. Recent spectroscopic studies show equivalent widths and profiles of the emission lines varying with a period of 620 s, and this certainly suggests an orbital interpretation (Steeghs 2003, Woudt & Warner 2003). Thus the weight of evidence certainly favors the orbital view.

## 5. WHY NO SUPERHUMPS?

This leaves unanswered the question of why ES Cet does not show superhumps. Stars can avoid superhumps by having too high a mass ratio ($q=M_2/M_1 \geq 0.4$ seems to forbid them), or too low an $\dot{M}$ ($\dot{M} \leq 10^{-9}$ $M_\odot$/yr, or $M_V > +8$, seems to forbid them). At $P_o = 620$ s with a degenerate secondary, ES Cet should have $M_2 \sim 0.1$ $M_\odot$ and therefore $q$ probably near 0.15 (see Figure 2 of Faulkner, Flannery & Warner 1972); so that solution probably does not work here.

The issue of $\dot{M}$ is more open, since we do not know the distance to ES Cet. In general, $\dot{M}$ and $M_V$ are correlated with the equivalent width of emission lines (Figure 6 of Patterson 1984); but this relation is not calibrated for helium CVs, so we now seek to remedy this.

There are just two distance-finding calibrators we could consider "primary": trigonometric parallax, and spectrophotometry of the mass-donor star. Among the AM CVns there are just three useful parallaxes (GP Com and V396 Hya, Thorstensen 2004; AM CVn, Dahn 2004) and no detected donors. A secondary method is a disk model fit to accurate spectrophotometry (continuum + lines) over a substantial wavelength interval. This method was applied to AM CVn and CR Boo (El-Khoury and Wickramasinghe 2000; Nasser, Solheim, & Semionoff 2001). Finally we could consider a standard-candle method: using the $M_V$'s of AM CVn and CR Boo to estimate those of other class members. This last method we would consider "tertiary". But it is known to work quite well for hydrogen-rich dwarf novae in eruption, and is a plausible consequence of the well-known thermostat resident in the accretion disks of dwarf novae (Warner 1987, Cannizzo 1998).

We collect the relevant information in Table 3. Most is self-explanatory, but a few points deserve mention. We cite references pertinent to the issues considered here, not necessarily the most complete or recent studies. Distance estimates from primary and secondary indicators are listed without comment, whereas those from a standard-candle method are less reliable and listed in parentheses. Of course, no distance is estimated from emission lines, since that is what we wish to establish. Mass ratios are estimated from the superhump period excess or the motion of spectral lines. We exclude from Table 3 one certain AM CVn member (KL Draconis) and two other possibles (V407 Vulpeculae and RXS J0806.3+1527) — basically because not enough information is yet available on them.

Figure 6 shows the correlation of the equivalent width of He I λ5876 with $M_V$. Very weak emission components are sometimes seen in the high states; the strengths are hard to measure, so we characterize them as "<1 Å". We also use information from individual stars in different states. Figure 6 shows the same general trend exhibited by hydrogen-rich CVs: strong emission when (intrinsically) faint, weak emission when bright. ES Cet has a line of 5 Å equivalent width, suggesting $M_V \sim 9.4$ (and $d \sim 350$ pc) according to Figure 6. Assuming a bolometric correction similar to that of other AM CVn stars,[4] this suggests $\dot{M} \approx 10^{-10} \, M_\odot$/yr and therefore might explain why ES Cet does not superhump: because the accretion rate is too feeble.

However, this argument is only suggestive, since the data of Figure 6 are so sparse and since the commonality of ES Cet with the other stars is unproven. So we will adopt this as only a *nominal* distance. What other constraints are there on $\dot{M}$ ?

## 6. ULTRAVIOLET AND X–RAY OBSERVATIONS: HOW MUCH TOTAL FLUX?

For a star as blue as ES Cet, the V magnitude greatly underestimates the bolometric flux. The star was detected in the first release of GALEX photometric magnitudes (Schiminovich 2004). The FUV (centered at 1520 Å) and NUV (2310 Å) magnitudes were found to be 15.25 and 15.7(±0.1) respectively on the AB magnitude scale. This corresponds to a dereddened continuum distribution of $F_\nu \sim \nu^{1.3}$, which is essentially the bluest known CV.

The flux distribution is shown in Figure 7, where we include also the hard X-ray flux detected by CHANDRA (Strohmayer 2004) and the soft X-ray flux marginally detected by ROSAT (Voges et al. 1999). The former is clearly not significant, comprising only 0.1% of the total flux. There is little direct spectral information in the 7 detected photons of the latter observation; but the spectrum must be very soft, in order not to conflict with the very low flux seen above 0.5 keV by CHANDRA. We experimented with blackbody fits to the ROSAT data, for the interstellar absorption expected along that line of sight ($N_H = 2 \times 10^{20}$ cm$^{-2}$, Schlegel et al. 1998). For the range of temperatures appropriate to "one-component" fits (including also the optical-UV flux), we find fluxes near the diamond in Figure 7.

We can learn more by studying the flux distributions of model stars. The vast majority of CVs do not have appreciable flux above the Lyman limit, due to H opacity in the disk's outer layers; the same applies, *mutatis mutandis*, to a hot helium disk, due primarily to the He II edges at 228 and 912 Å. But for ES Cet the blue optical-UV color implies very high temperature, enough to erode or maybe even destroy the edges. The helium-rich white dwarf models of Wesemael (1981) explore this in detail. In Figure 7 we superpose the model flux distributions at $T$=100000 and 150000 K. These appear to bracket the observations fairly well, and we conclude that a fairly good fit to all the data can be found with one component at $T$=130±15 kK, The total flux in such a component is $8(\pm 4) \times 10^{-10}$ erg/cm$^2$/s, which corresponds to a luminosity $L = 8(\pm 4) \times 10^{33} \, d_{300}^2$ erg/s, where $d_{300}$ is the distance in units of 300 pc.

This is essentially the maximum-luminosity solution for ES Cet. A two-component fit is possible, but produces less total flux. The ROSAT data can be fit by lower temperatures, but only by overproducing optical/UV emission. The data can be fit by higher temperatures, but then the luminosity of the soft X-ray source is small, since most of it actually falls above 0.1

---

[4] This will turn out to be wrong. The distance estimate may or may not be reasonable, but we will see that the EUV emission (evinced by the He II lines) is much greater in ES Cet — implying a higher bolometric correction and therefore a higher $\dot{M}$.

keV, i.e. in the ROSAT passband.

Another constraint can come from the He II 4686 emission, whose great strength (80 Å equivalent width) testifies to a large supply of photons with energy above 54 eV. The observed luminosity in the line is $10^{30} d_{300}^2$ erg/s. Repeating the calculation in Eq. (6) of Patterson & Raymond (1985) for a *soft* X-ray source, we estimate an ionizing luminosity $L_{SX} \sim (1–10) \times 10^{33} d_{300}^2$ erg/s. This corresponds to an unabsorbed flux roughly shown by the cross in Figure 7, where its location suggests that this is probably the X-ray source (barely) seen by ROSAT.

An alternative version of this argument, from the full range of Wesemael's helium-atmosphere models, can be made as follows. We have mined his tables and figures for the dependences of the following parameters on $T_{eff}$: (a) the bolometric correction relative to the optical-UV flux; and (b) the bolometric correction relative to the $V$ flux. We mean by "bolometric correction" the "correction in magnitudes for flux outside the relatively accessible 912–10000 Å and 5000–6000 Å bands". These dependences are shown in Figure 8.

For temperatures below ~$10^5$ K, Figure 8 shows quite modest corrections. Even at $10^5$ K, the optical/UV sum only undercounts the flux by 2.5 magnitudes, or a factor of 10. Can the temperature be as high as $3 \times 10^5$ K, in which case the undercount is much more serious — a factor of 100? Well, no, this is not feasible. So high a temperature, if required also to reproduce the optical-UV data, would produce a very luminous soft X-ray source — in contrast to the very weak source seen by ROSAT. If not required to reproduce the optical-UV flux, such a temperature is feasible but cannot represent much luminosity (since it does not leak appreciably into the UV or hard X-ray, and produces a pretty wimpy ROSAT detection). For the one component solution near 130 kK, the bolometric corrections shown in Figure 8 yield a total luminosity of $1.0(\pm 0.2) \times 10^{34} d_{300}^2$ erg/cm$^2$/s. Multiple-component solutions exist, of course, but are of lower luminosity.[5] Solutions of *much* lower luminosity are excluded, because they cannot power the observed He II emission.

We shall adopt as a final answer $L = 1.0^{+0.2}_{-0.4} \times 10^{34} d_{300}^2$ erg/s.

In pure disk accretion, the accretion rate $\dot{M}$ should be

$$\dot{M} = 2LR_1/GM_1$$

where $M_1$ and $R_1$ are the mass and radius of the accreting white dwarf. This implies $\dot{M}$ in the range $(1.3–2.6) \times 10^{-9}$ $(M_1/0.7\ M_\odot)^{-1.7} d_{300}^2 M_\odot$/yr, or $(2–4) \times 10^{-9}$ $(M_1/0.7\ M_\odot)^{-1.7} M_\odot$/yr at the nominal 350 pc distance. Here we have assumed $R_1 \propto M_1^{-0.7}$, appropriate for a white dwarf of moderate mass.

---

[5] There might also be a worry about variability, since the observations relevant to Figure 7 were obtained at widely different times. But ES Cet deeply impressed us with its constancy in visual light. Although we cannot rule out changes of up to 0.1 mag, we essentially saw the same mean brightness on every night of the campaign. Nor is there any historical evidence for brightness change. Indeed, apart from the strictly periodic variation, it is the most constant CV we have ever observed.

## 7. ACCRETION RATES IN THE AM CVN CLASS

We repeated this analysis for other members of the AM CVn class. These are easier, because the continuum slopes are more modest, implying small bolometric corrections (even for the hottest, probably AM CVn, the He II edges largely suppress emission at EUV wavelengths). After integrating the observed 0.1–1.0 μ fluxes and applying the (small) bolometric corrections, or estimating for the stars without published ultraviolet data, we used our distance estimate to estimate bolometric luminosities. We then calculated mean accretion rates from the assumption of pure disk accretion onto a 0.7 $M_\odot$ white dwarf. These are plotted versus $P_o$ in the lower frame of Figure 9. For the three dwarf novae we took care to average over a cycle of eruption and quiescence. The error shown includes uncertainty in bolometric flux, $M_1$ (0.5–1.0 $M_\odot$), and distance (except for ES Cet).

How do these $\dot{M}$'s compare with the theory of accretion driven by GR? Well, $\dot{M}$ should scale roughly as $P_o^{-5}$ — but also depends on the secondary's mass–radius relation. Reckoned in solar units, three mass–radius relations provide a convenient benchmark for discussion:

(a) A cold ("Zapolsky–Salpeter") white dwarf, supported purely by degeneracy pressure. A slight simplification from Eq. (5) of Nelemans et al. (2001), and valid over the relevant domain ($M_2$=0.01–0.30), gives

$$R_2 = 0.0152 \, M_2^{-0.22}. \qquad (2a, ZS)$$

(b, c) An evolved helium star, which begins mass transfer before the end of nuclear burning. This possibility was anticipated in the earliest discussions of AM CVn (Faulkner, Flannery, & Warner 1972). More recent models which explicitly calculate the mass–radius relation, accounting for GR and the secondary's cooling, have been published by Savonije, de Kool, & Van den Heuvel (1986, hereafter SKH) and Tutukov & Fedorova (1989, hereafter TF). These give mass–radius relations

$$R_2 = 0.029 \, M_2^{-0.19} \qquad (2b, SKH)$$

and

$$R_2 = 0.043 \, M_2^{-0.06} \qquad (2c, TF)$$

Combined with Kepler's Third Law and Paczynski's (1971) approximation for the Roche-lobe size in a close binary, the corresponding mass–period relations are

$$M_2 = 0.033 \, P_{1000}^{-1.21} \qquad (3a, ZS)$$
$$M_2 = 0.093 \, P_{1000}^{-1.27}, \qquad (3b, SKH)$$

$$M_2 = 0.119\, P_{1000}^{-1.70}, \qquad (3c,\ \text{TF})$$

where $P_{1000} = P_o/1000$ s.

These differences in mass–radius and mass–period make a significant difference in the expected $\dot{M}$. From the angular-momentum loss prescription in GR (Landau & Lifshitz 1959), a few pages of algebra yield accretion rates[6]

$$\dot{M}_2 = 3.8 \times 10^{-10}\ M_\odot/\text{yr}\ M_1\ \text{M}^{-1/3}\ P_{1000}^{-5.10}\ (0.72 - q)^{-1} \qquad (4a)$$

$$\dot{M}_2 = 3.2 \times 10^{-9}\ M_\odot/\text{yr}\ M_1\ \text{M}^{-1/3}\ P_{1000}^{-5.21}\ (0.73 - q)^{-1} \qquad (4b)$$

$$\dot{M}_2 = 5.2 \times 10^{-9}\ M_\odot/\text{yr}\ M_1\ \text{M}^{-1/3}\ P_{1000}^{-6.07}\ (0.80 - q)^{-1} \qquad (4c)$$

for these mass–radius relations. These theoretical runs of $\dot{M}(P)$, assuming a 0.7 $M_\odot$ primary, are the solid lines in the lower frame of Figure 9.

How do the points compare with the solid-line predictions in Figure 9? In general, the points track along a $P_o^{-5}$ curve pretty well, as predicted. Most of them could be consistent with either a cold (ZS) white-dwarf secondary, or a larger SKH/TF secondary. ES Cet seems to be consistent only with a ZS secondary; but that error bar (only) does not include distance uncertainty, which is considerable since the star is arguably *sui generis*. At 1 kpc, which is possible, the star would jump up to the SKH prediction.

Three other constraints are significant. *First*, some AM CVn stars suffer disk instabilities, and others don't. The theory for this has been developed by Smak (1983) and Tsugawa & Osaki (1997); the dashed lines in the lower frame of Figure 9 show the results of Tsugawa & Osaki. Below $\dot{M} \sim 10^{-11}$ (depending slightly on $M_1$), disks are too cool to ionize helium anywhere, and accrete more or less steadily. Above the upper dashed line, disks are too hot everywhere to suffer the ionization instability. For intermediate $\dot{M}$, disks cycle between cold and hot states, and the stars consequently show dwarf-nova eruptions. All the AM CVn binaries obey these predictions (the three dwarf novae are CR Boo, CP Eri, and V803 Cen). So far, so good.

*Second*, we have independent evidence regarding the masses of the secondaries, via the superhumps which tend to dominate their light curves. At least for H-rich CVs, there is a good correlation between mass ratio and the observed fractional superhump period excess $\varepsilon = (P_{sh} - P_o)/P_o$, namely $\varepsilon = 0.22(2)q$ (Patterson 2001). So we know four mass ratios, plus two others (GP Com and V396 Hya) based on the small orbital wiggles in the emission lines (Morales-Rueda et al. 2003, Ruiz et al. 2001). These estimates are given in Table 3, and shown in the upper frame of Figure 9. The curves show predictions for two versions of mass–radius. Again the points are roughly between the theory curves, though at short period there is a decided preference for SKH.

---

[6] See Paczynski (1967) and Nelemans et al. (2001) for fuller discussion. The key point is that $\dot{M}$ scales as $(m_2)^2$ at a fixed period, so the large spread in $m_2(P)$ shown in Eqs. (3) causes a large spread in $\dot{M}$.

And *third*, it appears from studies of H-rich CVs that there is a magic number (roughly $10^{-9}$ $M_\odot$/yr but possibly defined by the upper dashed line in Figure 9), which determines the presence or absence of superhumps.[7] How does this expectation fare in Figure 9? Leaving out ES Cet, the two stars of highest $\dot{M}$ superhump; the two stars of lowest $\dot{M}$ don't; and the three stars of intermediate $\dot{M}$ basically do, but only when $\dot{M}$ is higher (superoutburst). This is quite consistent with what we have learned about superhumps in CVs, except for the embarrassing noncompliance of ES Cet. In Sec. 5 we speculated that insufficient $\dot{M}$ in ES Cet might be the culprit; but after accounting for the observed and inferred EUV radiation, we see that superhump absence is still unexplained.

A possible clue to that absence can be found by estimating the size of the emitting region. In order to produce the correct flux at the ~130 kK temperature suggested by Figure 7, the luminous area should have a radius of $4.5\times10^8 d_{300}^{1.0}$ cm, assuming $A=\pi\times R^2$ (appropriate for a disk or white dwarf). This could represent a fairly massive (1.1 $M_\odot$) white dwarf, or a large spot on a white dwarf. It is not so easily reconciled with a disk, which should have $R\sim14\times10^8$ cm. This recalls the "direct impact" theory of Marsh & Steeghs (2002, hereafter MS), who calculated that the mass-transfer stream should strike the white dwarf rather than form a disk — for a white dwarf binary with $P$=9.5 minutes (V407 Vul). That would explain it: no superhumps because there's no disk. We repeated the MS calculation for the slightly longer period (10.3 versus 9.5 minutes), and verified that there is no great sensitivity to period. There is still a wide swath of $M_1$–$M_2$ space (shown in Figure 2 of MS) where direct impact is expected, as long as $M_1$<0.8 $M_\odot$ (more massive primaries are smaller than the stream's impact parameter, and are therefore missed).

This seems pretty plausible to us, and we take the small emission area to be an empirical clue that it is the correct explanation. The data of Figure 9 seem compliant in all other respects with the simple theory invoking GR only as the driver of evolution: the accretion rates are about right and show the correct scaling with $P_o$; the eruptive stars are just those in the unstable-disk regime; and the superhump characteristics are normal except for ES Cet, which may well not have a disk. The House of Can Ven appears to be in satisfactory order.

## 8. SUMMARY AND OUTLOOK

1. ES Cet's magnitude remained roughly constant at $V$=17.0 throughout our three years of photometry, with little flickering. The powerful periodic signal discovered by WW was detected on every night at 139.3 c/d, along with its first three harmonics.

2. The signal's full amplitude was 0.12 mag, with little variability. The phase remained measurably constant over the 3 year baseline.

---

[7] It is more correct to associate this with viscosity, since that is what enlarges the disk and gives access to the 3:1 resonance which evidently drives the superhump (eccentric) instability. But in steady accretion, viscosity and $\dot{M}$ are directly related; and since we "observe" $\dot{M}$ rather than viscosity, it is better to cite the dependence on $\dot{M}$.

3. Most other stars in the AM CVn class are dominated by a superhump variation; we studied the ES Cet time series to find the telltale signatures of superhumps, and mostly failed to find them. The power spectra did show some "blue bumps" ~12 c/d displaced from the main signal, but the bumps were weak and not displaced by any precise amount. With a strictly constant phase and no repeatable fine structure (to our limits of measurement), the main signal most likely denotes the true orbital period of the binary.

4. We present an empirical correlation between $M_V$ and the equivalent width of He I $\lambda 5876$ emission, which resembles the well-known relation for hydrogen-rich CVs. We use this to extract a nominal distance estimate of 350 pc to ES Cet, although a better estimate of distance is sorely needed.

5. The ultraviolet magnitudes of ES Cet establish it to be a very blue star — essentially bluer than any known CV. Available constraints from X-ray observation (ROSAT detection and a strong CHANDRA limit at 0.5 keV), plus the great strength of He II 4686 emission, establish that most of the flux is emitted at EUV wavelengths. A one-component fit to the data gives a temperature of 130±10 kK, and a luminosity of $1.0^{+0.2}_{-0.4} \times 10^{34} d^2_{300}$ erg/s.

6. For accretion onto a 0.7 $M_\odot$ white dwarf, this yields $\dot{M}=(2-4)\times 10^{-9} M_\odot$/yr at our nominal 350 pc distance. We compare this to average $\dot{M}$ s estimated for other AM CVn stars, and to the rates expected if mass transfer is solely due to GR losses. The observed rates seem to follow the predicted $\dot{M} \sim P_o^{-5}$ scaling, and to agree in normalization. The stars also show the properties normally associated with those accretion rates: presence or absence of outbursts, presence or absence of superhumps. The one exception is ES Cet, which should show superhumps and evidently doesn't.

7. We estimate a size for the emitting region in ES Cet: $4.5\times 10^8 d^{1.0}_{300}$ cm. This seems too low to be the accretion disk, but could be the white dwarf, or a large spot on the white dwarf. This provides some evidence supporting the idea that in the ultracompact binaries, the mass-transfer stream directly impacts the white dwarf, rather than forming an accretion disk. It would also explain why ES Cet fails to show superhumps: because there is no disk.

8. The placement of stars in Figure 9 includes the uncertainty in bolometric flux, an assumed uncertainty in $M_1$ (0.5–1.0 $M_\odot$), and the uncertainty in distance — except that no distance uncertainty is assigned to ES Cet since we regard it as *sui generis*. Since accretion rates scale as $d^2 M_1^{-1.7}$, more accurate constraints on $M_1$ and distance in all of these stars would go a long way towards refining the test shown in Figure 9.

We gratefully acknowledge discussions and communication of unpublished data from Conard Dahn, John Thorstensen, and Tom Marsh; and observational assistance from Jonathan Kemp and Eve Armstrong. We relied on financial support from the NSF (AST–00–98254) and STSCI (NST 90–9459.04A).

TABLE 1
LOG OF OBSERVATIONS

| Start: HJD 2,452,000+ | Duration of Observation (d) | Start: HJD 2,452,000+ | Duration of Observation (d) |
|---|---|---|---|
| 201.3893 | 0.1805 | 553.5397 | 0.0999 |
| 202.3578 | 0.1671 | 553.8534 | 0.1159 |
| 203.3739 | 0.1488 | 554.3096 | 0.0841 |
| 204.5488 | 0.0756 | 554.8621 | 0.0994 |
| 292.5686 | 0.1259 | 555.3628 | 0.0458 |
| 293.5717 | 0.1329 | 555.5541 | 0.0319 |
| 318.5915 | 0.0484 | 556.3182 | 0.1422 |
| 320.5821 | 0.0561 | 556.5616 | 0.0398 |
| 321.2701 | 0.0423 | 556.7612 | 0.2215 |
| 322.2753 | 0.0458 | 558.5422 | 0.0443 |
| 323.2645 | 0.0410 | 577.3226 | 0.0433 |
| 324.2658 | 0.0319 | 604.3599 | 0.0499 |
| 327.2612 | 0.0469 | 608.3449 | 0.0372 |
| 430.6509 | 0.0336 | 615.5669 | 0.0366 |
| 434.6498 | 0.0379 | 626.3345 | 0.0129 |
| 436.6415 | 0.0315 | 634.2849 | 0.0356 |
| 514.4789 | 0.1550 | 636.5787 | 0.0922 |
| 515.4630 | 0.0745 | 639.2979 | 0.0324 |
| 518.4776 | 0.0278 | 673.2670 | 0.0323 |
| 544.8311 | 0.1893 | 798.6440 | 0.0384 |
| 545.8119 | 0.2059 | 800.6438 | 0.0369 |
| 546.8067 | 0.2085 | 872.6075 | 0.0538 |
| 548.8090 | 0.0906 | 874.6491 | 0.0333 |
| 550.7749 | 0.2309 | 904.5818 | 0.0611 |
| 551.3989 | 0.2477 | 918.5173 | 0.0437 |
| 552.4585 | 0.0308 | | |

TABLE 2
TIMES OF 620 s MINIMUM LIGHT

| HJD 2,452,000+ | | | | |
|---|---|---|---|---|
| 201.3941 | 323.2686 | 545.8124 | 556.5658 | 639.3037 |
| 202.3632 | 324.2662 | 546.8103 | 556.7667 | 673.2720 |
| 203.3755 | 327.2668 | 548.8131 | 558.5471 | 798.6492 |
| 204.5527 | 430.6570 | 550.7800 | 577.3257 | 800.6450 |
| 292.5738 | 434.6556 | 551.4044 | 604.3666 | 872.6081 |
| 293.5786 | 436.6438 | 553.5437 | 608.3504 | 874.6541 |
| 318.5953 | 514.4790 | 553.8596 | 615.5719 | 904.5878 |
| 320.5837 | 515.4697 | 554.3118 | 626.3395 | 918.5211 |
| 321.2732 | 518.4846 | 554.8645 | 634.2861 | 1152.6506 |
| 322.2780 | 544.8362 | 556.3216 | 636.5832 | 1165.6437 |

NOTE. – The typical error for timing minimum light is ± 0.0002 days.

TABLE 3
PROPERTIES OF AM CVN STARS

| Star | State | $V$ | $M_V$ | EW(5876) (Å) | $d$ (pc) | $q=M_2/M_1$ | $F_X/F_V$ | References |
|---|---|---|---|---|---|---|---|---|
| GP Com | Low | 15.7 | 11.5 | 35 | 70 | 0.020(6) | 0.7 | Nather et al. 1981, Williams 1983, Thorstensen 2003 |
| AM CVn | High | 14.1 | 7.2 | abs | 240 | 0.101(8) | 0.06 | El-Khoury & Wickramasinghe 2000, Smak 1975, Dahn 2004 |
| HP Lib | High | 13.6 | | abs | (330) | 0.072(6) | <0.2 | O'Donoghue et al. 1994 |
| V396 Hya | Low | 17.7 | 13.0 | 60 | 80 | 0.022(6) | <4 | Ruiz et al. 2001, Thorstensen 2004 |
| CP Eri | High | 16.5 | | abs | (800) | 0.040(4) | | Abbott et al. 1992 |
| CP Eri | Low | 19.7 | | 10 | | | <4 | |
| CR Boo | High | 13.4 | <6.4 | abs | >250 | 0.060(5) | 0.3 | El-Khoury & Wickramasinghe 2000, Wood et al. 1987, Patterson et al. 1997, Dahn 2004 |
| CR Boo | Low | 16.7 | <9.7 | 4 | | | | |
| CR Boo | Low | 17.2 | <10.2 | 8 | | | | |
| V803 Cen | High | 13.4 | | abs | (250) | | <0.5 | O'Donoghue et al. 1987, O'Donoghue & Kilkenny 1989, Patterson et al. 2000 |
| V803 Cen | Mid | 14.5 | | abs | | | | |
| V803 Cen | Low | 17.0 | | 6 | | | | |
| ES Cet | Low? | 17.0 | | 5 | | | 0.1 | Woudt & Warner 2003, Thorstensen 2004 |

NOTE. — $F_X/F_v$ denotes $F$(0.1–2.5 keV)/$F$(5000–6000 Å), with an estimated correction for absorption. Most CVs are hard X-ray sources, so that correction assumes a hard spectrum.

# FIGURE CAPTIONS

FIGURE 1. — Light curve of ES Cet obtained on 2002 September 28, at 10 s time resolution. Like most data discussed in this paper, this was obtained through a broad bandpass, centered near 5500 Å. We estimate a mean *V* of 17.0, although there could be a zero-point error of up to 0.15 mag.

FIGURE 2. — Power spectra of a seven-night light curve in 2002 October, with significant features marked with their frequencies in cycles/day. These show the powerful fundamental and the first three harmonics.

FIGURE 3. — Mean light curves of the 620 signal in 2001 October, 2002 September, and 2002 October. The signal is remarkably stable in period, amplitude, and waveform — from month to month, and even from night to night.

FIGURE 4. — O–C diagram of the 620 s minima, with respect to the test ephemeris of Eq. (1). The straight line of zero slope indicates no discernible period change, corresponding to $|\dot{P}|<1.5\times10^{-11}$.

FIGURE 5. — "Cleaned" power spectra of 2002 September and October (upper and lower panels respectively). The fundamental and harmonics have been subtracted from the time series, and then re-inserted into the power spectra, to draw attention to the "blue bump" flanking the main signals. This is typically displaced by ~12 cycles/day, but we found no convincing pattern in the precise frequencies.

FIGURE 6. — Equivalent width of He I λ5876 emission versus $M_V$, for AM CVn stars. The correlation found (strong emission when faint, weak emission when bright) mimics that found in H-rich CVs.

FIGURE 7. — Observed fluxes in ES Cet, corrected for reddening. The star shows a very blue optical-UV continuum, weak emission in the ROSAT PSPC (0.1–2.0 keV) band, and very weak emission in the CHANDRA (0.5–10 keV) band. The very low CHANDRA counts in 0.5–2.0 keV establish that ROSAT observed a *soft* X-ray source, and fits of that data to a soft source gave fluxes near the diamond. The cross estimates the flux above 54 eV inferred from the strength of the He II 4686 emission. The CHANDRA observation is represented by a 0.8 keV blackbody (Strohmayer 2004), and the two theory curves show the fluxes expected from helium-rich model atmospheres at 100 and 150 kK. A single temperature of 130±10 kK fits all the data.

FIGURE 8. — Theoretical "bolometric corrections" for hot helium-rich white dwarfs, from Wesemael (1981).

FIGURE 9. — *Lower frame*, empirical accretion rates versus $P_o$ for AM CVn stars, compared with three predictions based on GR and different mass–radius prescriptions for the secondary (solid lines: ZS, SKH, TF, see text). The boundaries in $\dot{M}$–$P_o$ space predicted by disk-instability theory are shown as dashed lines. *Upper frame*, trend of $q=M_2/M_1$ with $P_o$, compared with predictions from the three mass–radius relations.

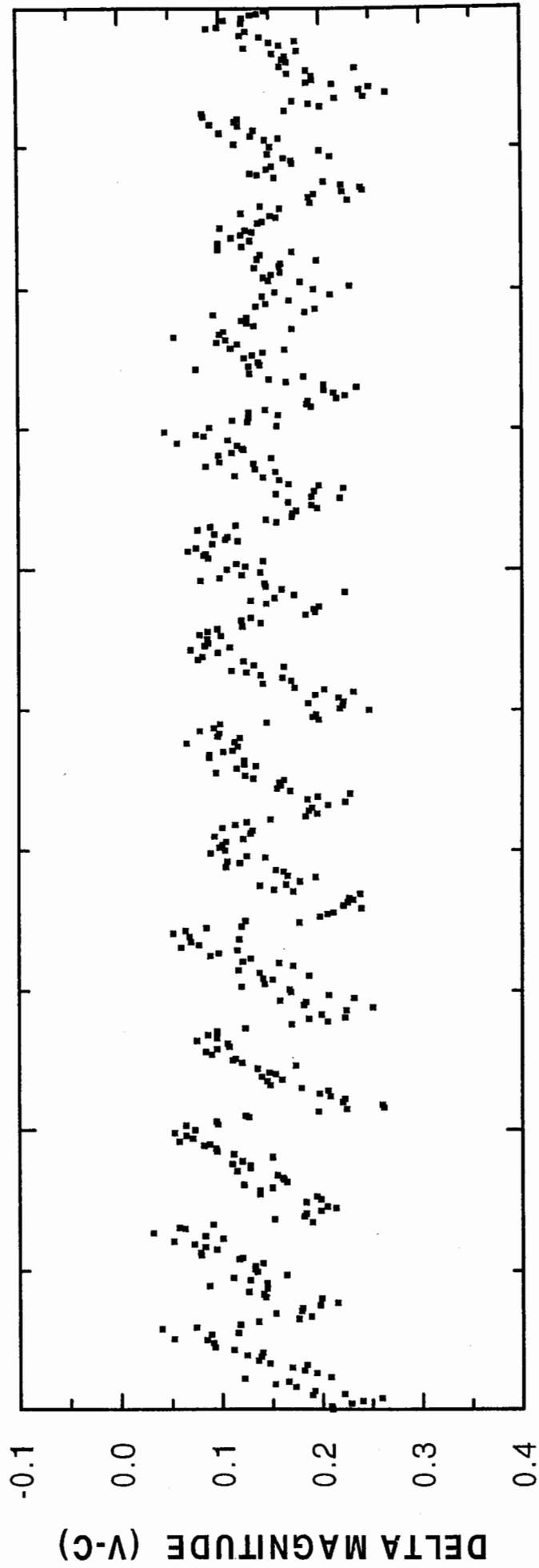

Fig 1

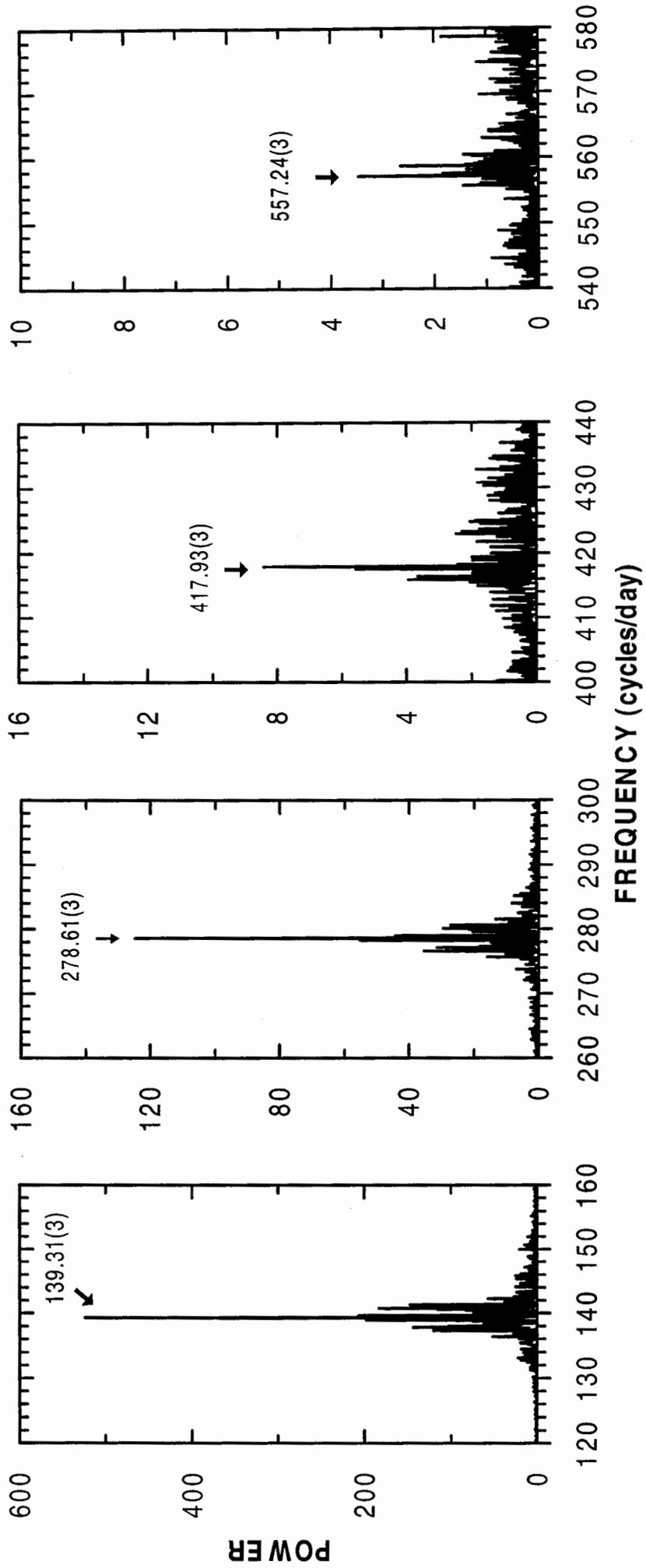

Fig 2

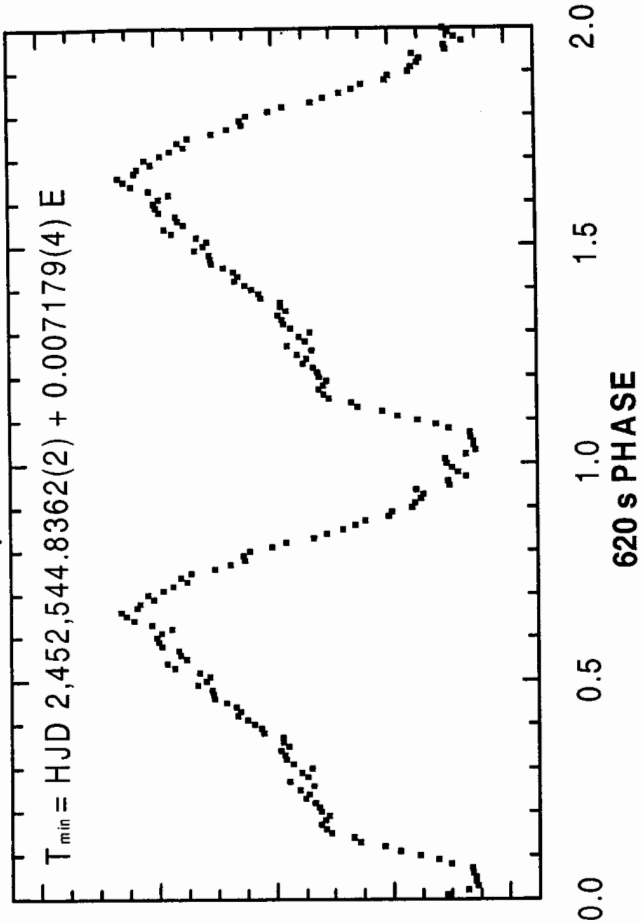
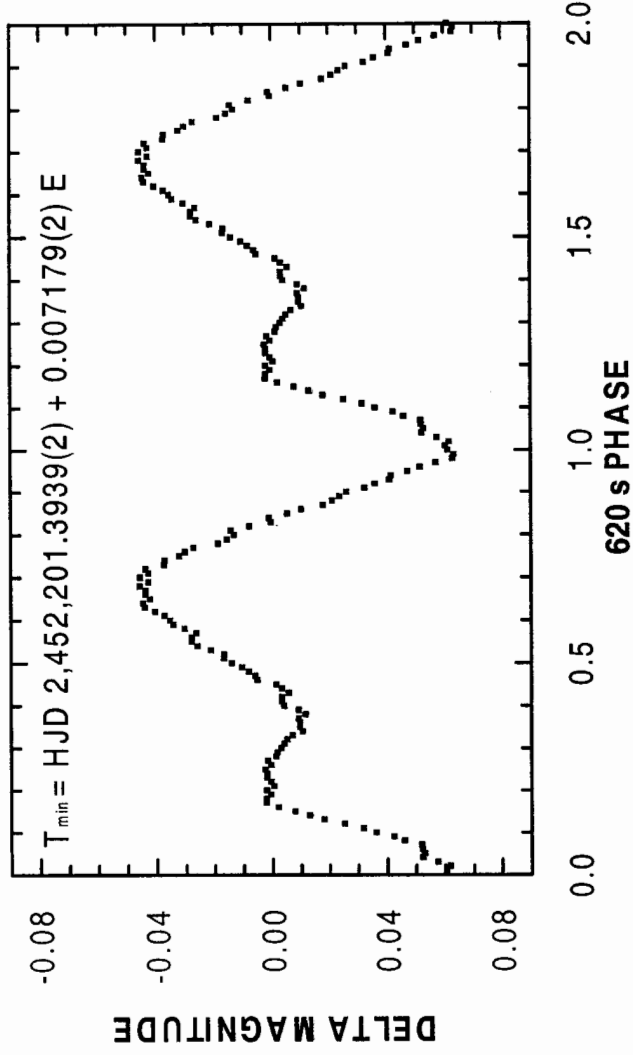
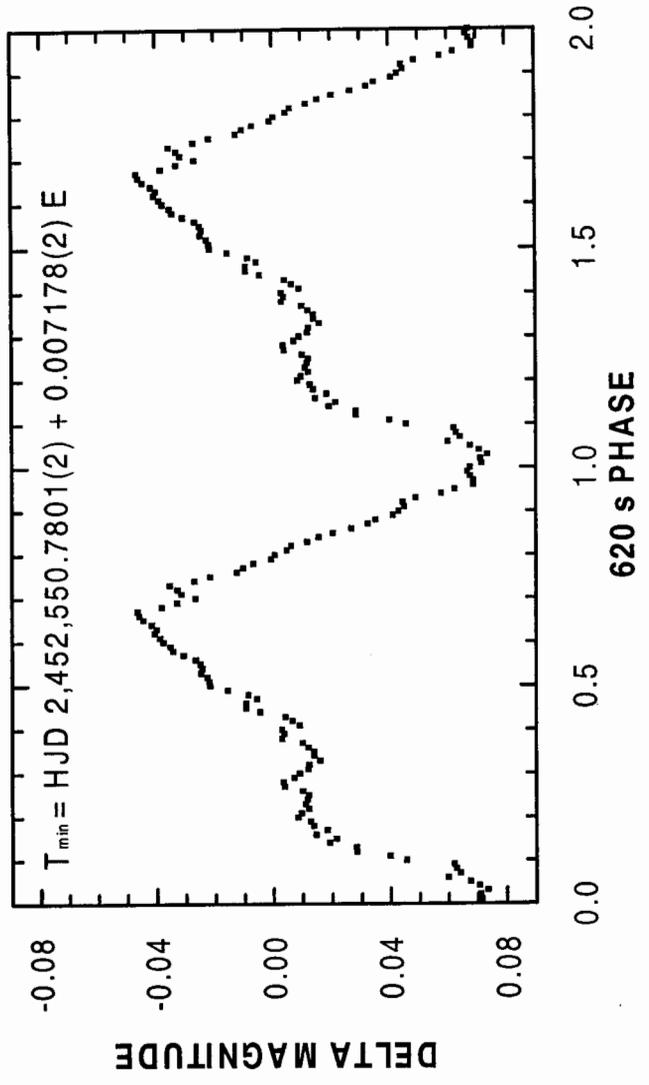

Fig 3

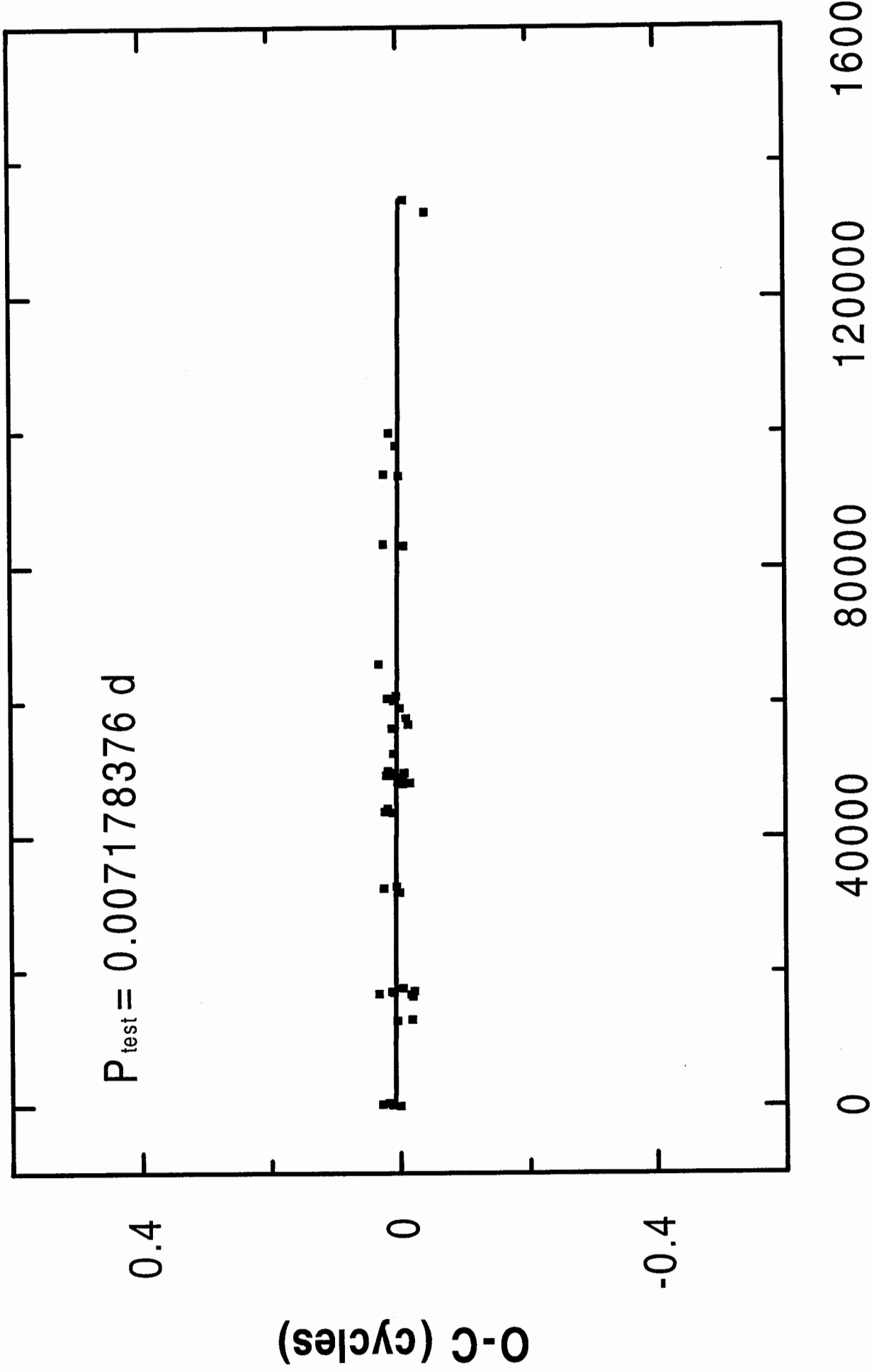

Fig 4

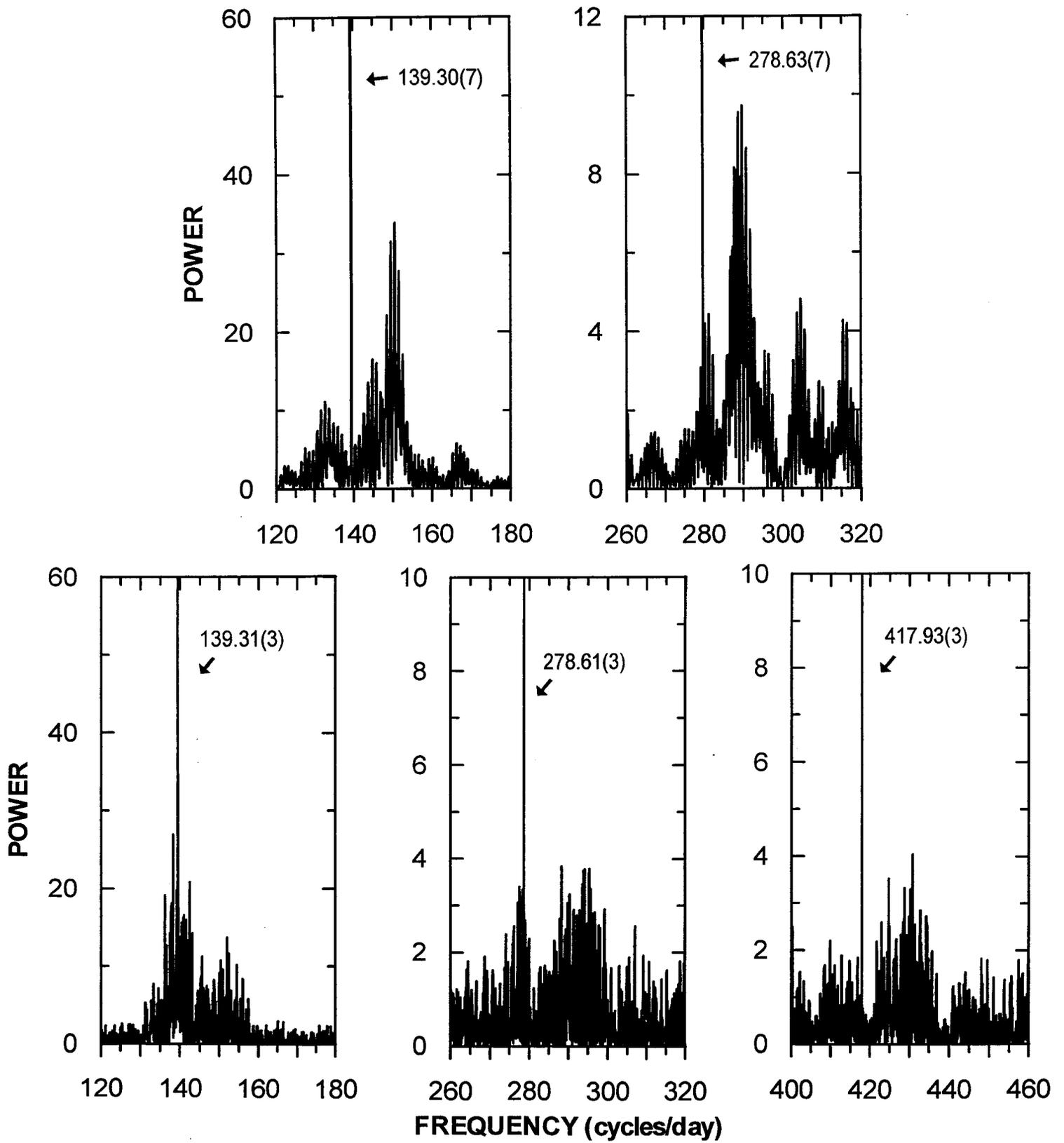

Fig 5

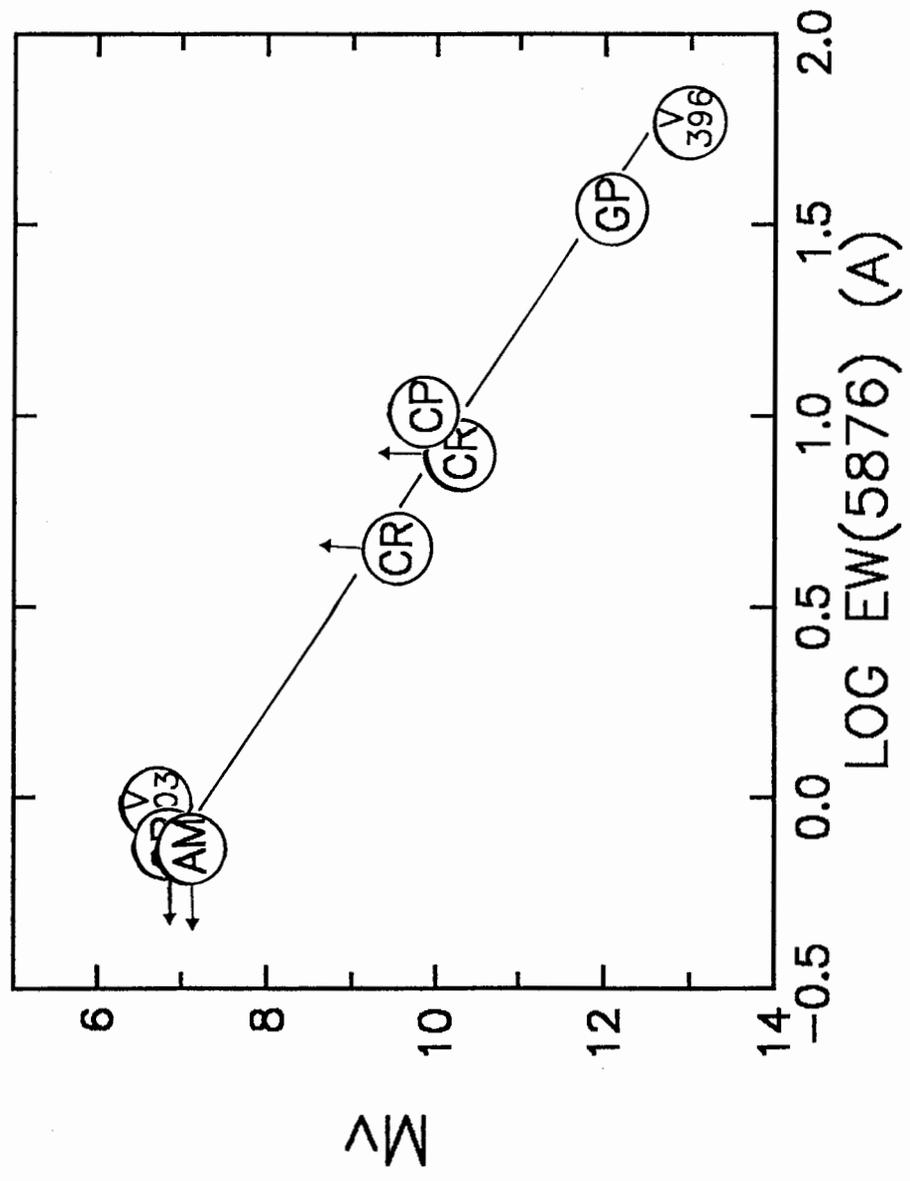

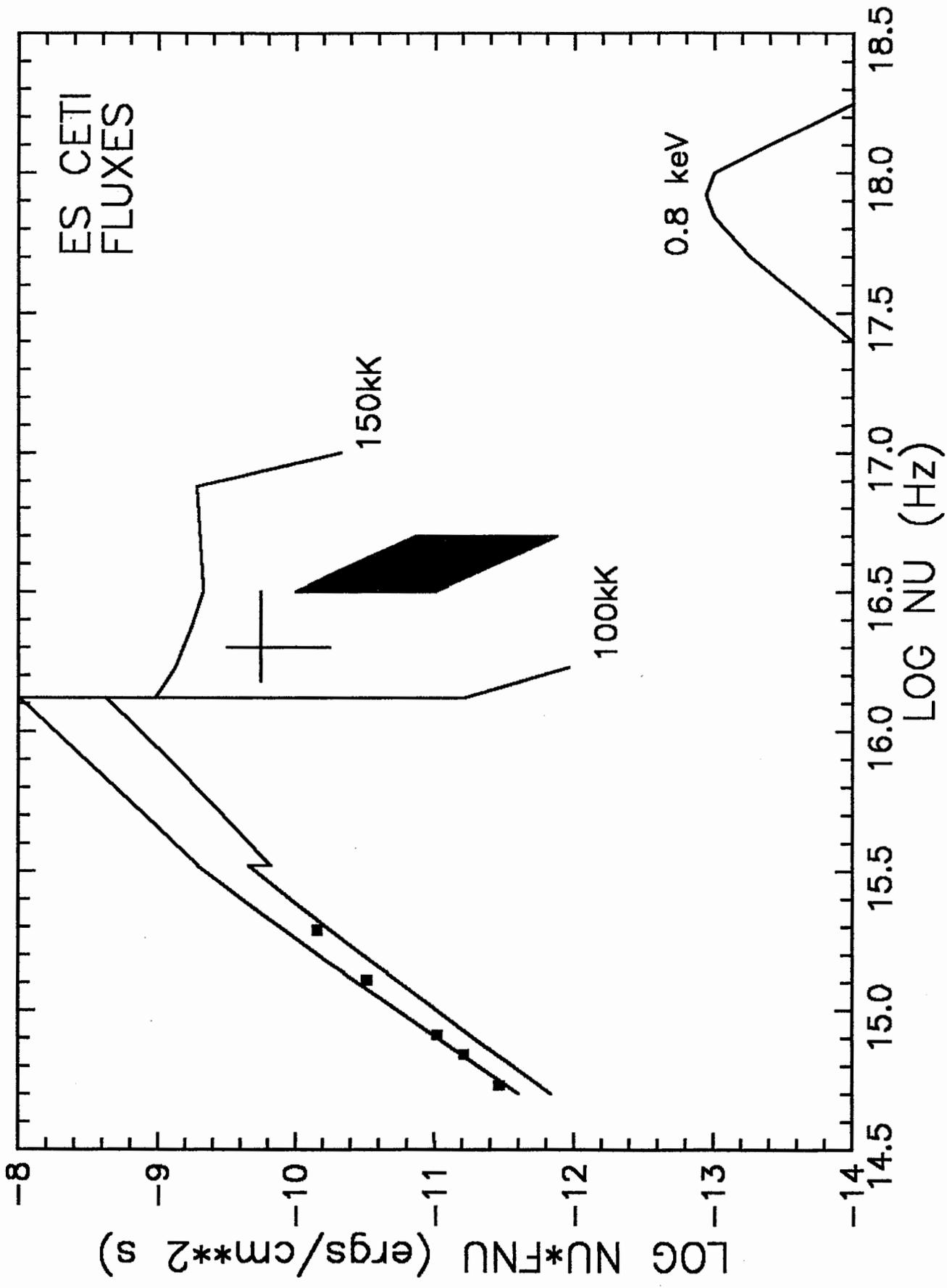

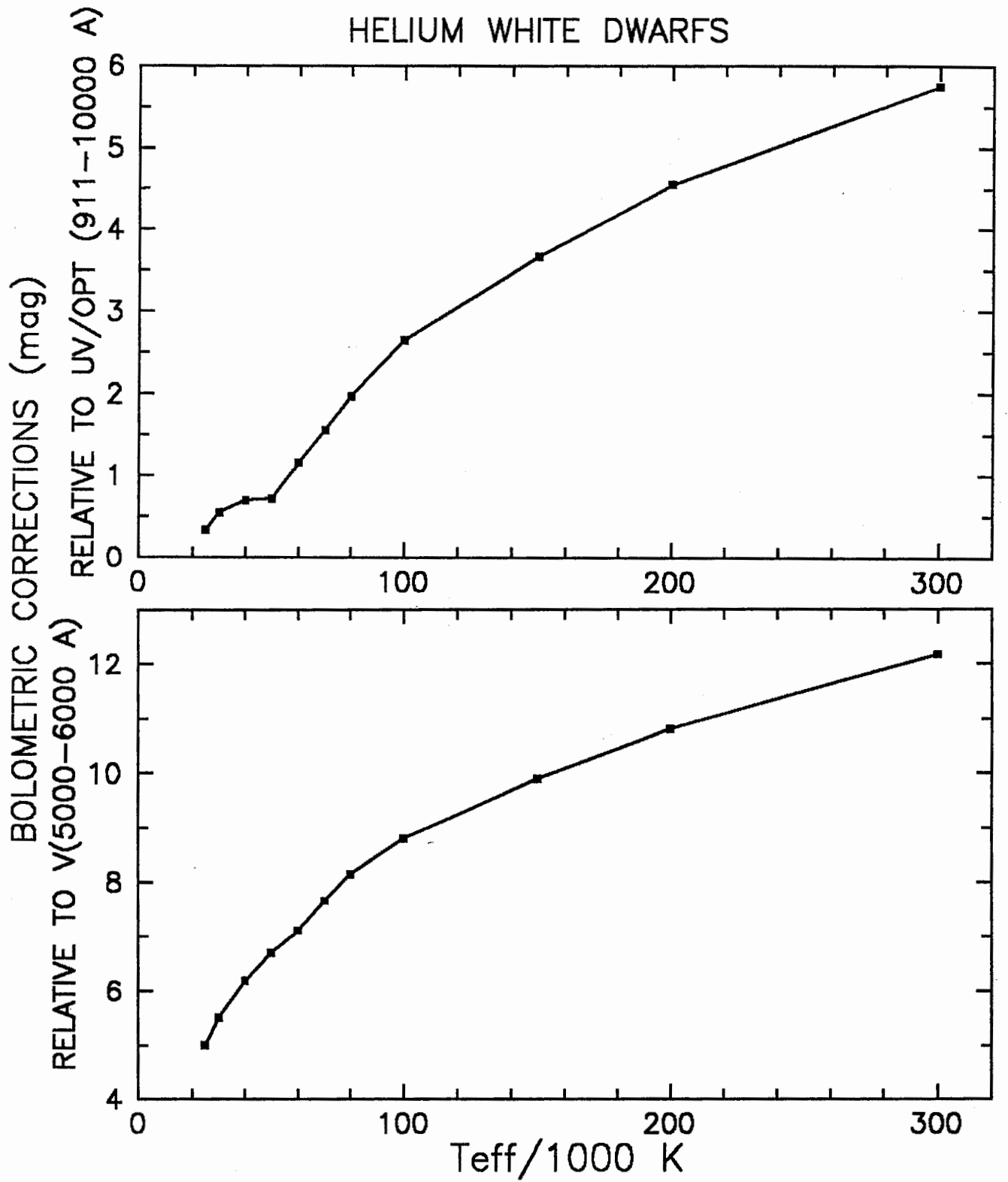

Fig 8

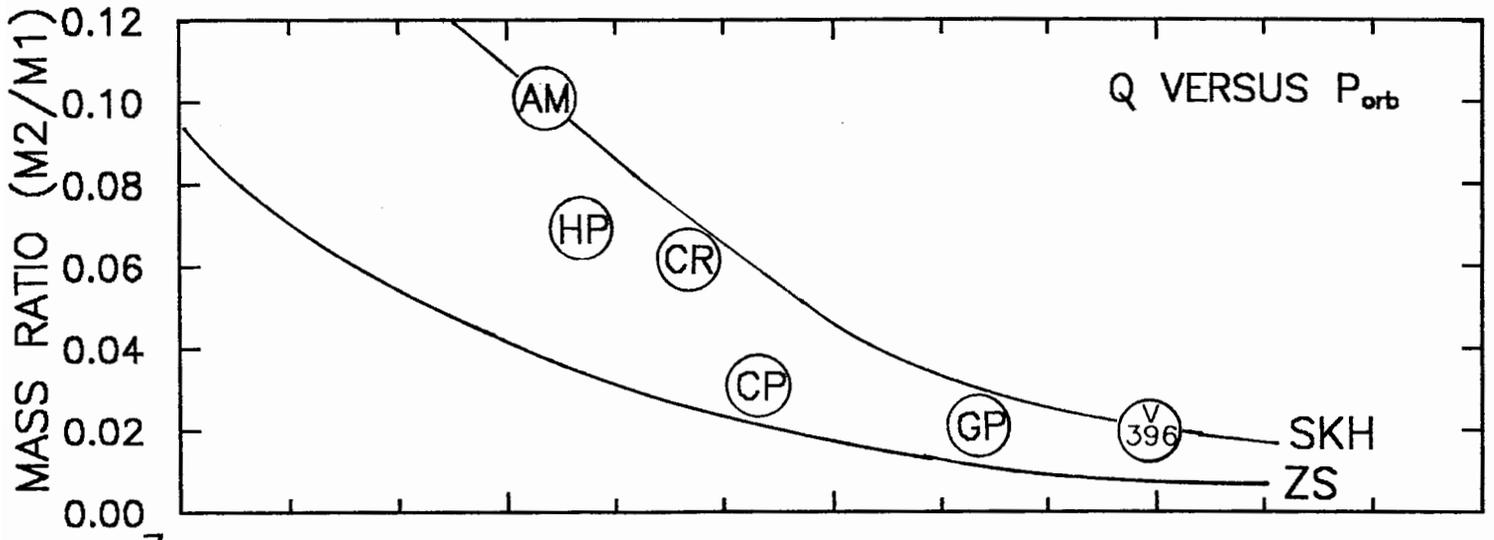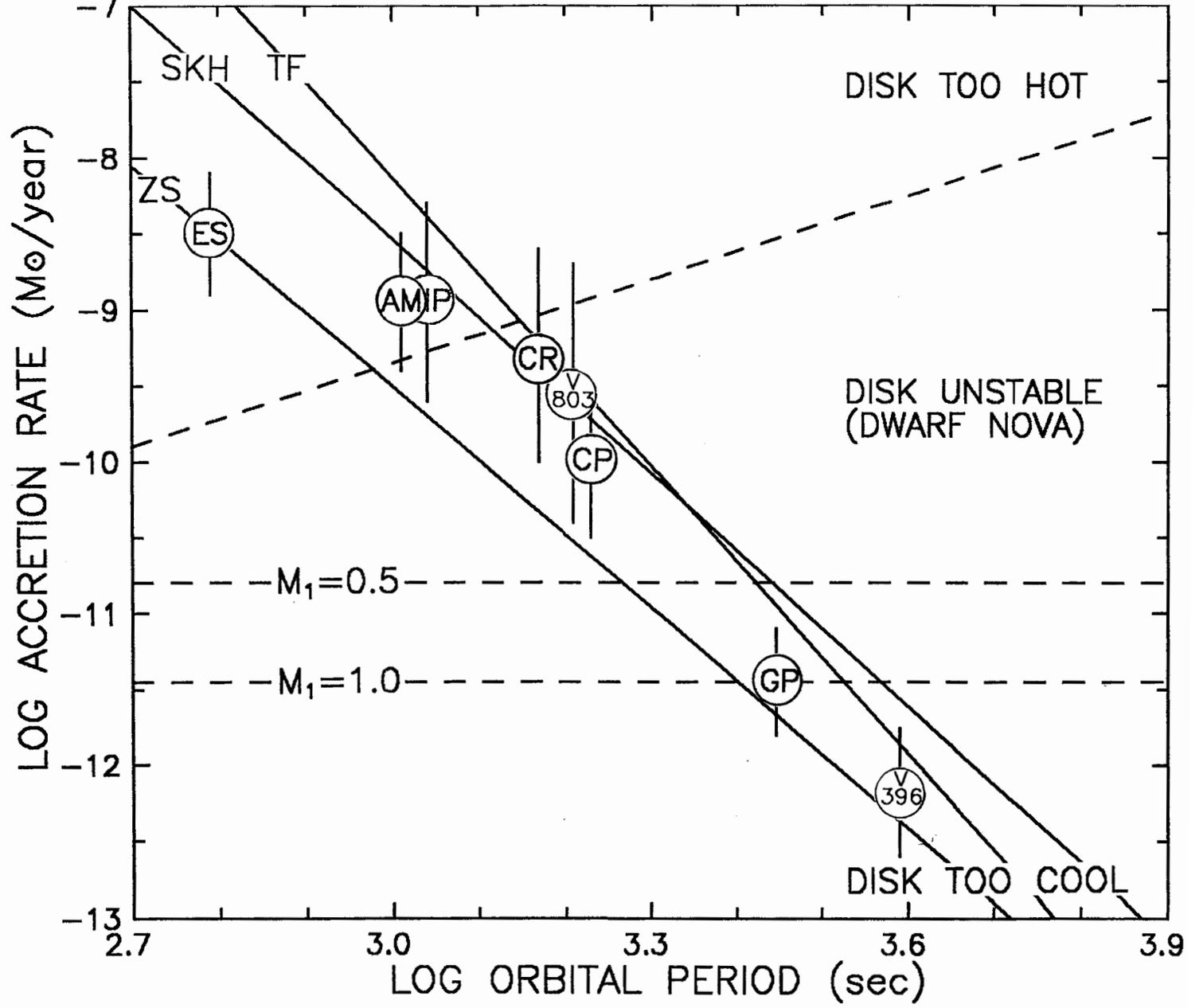

Fig 9